\documentclass[aip,apl,letterpaper,amsmath,amssymb,reprint]{revtex4-1}
\usepackage{graphicx}
\usepackage{microtype}

\begin{document}
\title{Tuning the perpendicular magnetic anisotropy, spin Hall switching current density and domain wall velocity by submonolayer insertion in Ta / CoFeB / MgO heterostructures}
\author{S. P. Bommanaboyena}
\affiliation{ Department of Materials Science, Technische Universit\"at Darmstadt, D-64287 Darmstadt, Germany}
\affiliation{Center for Spinelectronic Materials and Devices, Department of Physics, Bielefeld University, D-33501 Bielefeld, Germany}
\author{M. Meinert}
\affiliation{Center for Spinelectronic Materials and Devices, Department of Physics, Bielefeld University, D-33501 Bielefeld, Germany}
\email{meinert@physik.uni-bielefeld.de}

\date{\today}

\begin{abstract}
By submonolayer insertion of Au, Pt, or Pd into Ta / CoFeB / MgO / Ta heterostructures we tune the perpendicular magnetic anisotropy and the coercive field of the ferromagnetic layer. We demonstrate that this has a major influence on the spin Hall switching current density and its dependence on the external magnetic field. Despite a rather small effective spin Hall angle of $\theta_\mathrm{SH} \approx -0.07$, we obtain switching current densities as low as $2 \times 10^{10}$\,A/m$^2$ with a 2\,\AA{} Au interlayer. We find that the Dzyaloshinskii-Moriya interaction parameter $D$ is reduced with Au or Pd interlayers, and the perpendicular anisotropy field is reduced by an order of magnitude with the Pd interlayer. The dependence of the switching current density on the current pulse width is quantitatively explained with a domain wall nucleation and propagation model. Interface engineering is thus found to be a suitable route to tailor the current-induced magnetization switching properties of magnetic heterostructures.
\end{abstract}

\maketitle

The ability of spin-orbit torques, and in particular of the spin Hall effect\cite{Dyakonov1971, Hirsch1999, Hoffmann2013} (SHE), to generate  spin currents that can be used to manipulate the magnetization of an ultrathin magnetic  film has triggered very active research of the spintronics community on this novel field. The spin current generated through the SHE in a heavy metal film creates effective fields in an adjacent magnetic film which can be strong enough to excite magnetization dynamics and even magnetization switching.\cite{Liu2011} The pivotal quantity of the SHE is the spin Hall angle $\theta_\mathrm{SH} = j_\mathrm{s} / j$ describing the ratio of the spin current $j_\mathrm{s}$ and the orthogonal charge current $j$. Large values of the spin Hall angle were reported for the systems Ta / CoFeB ($\theta_\mathrm{SH} = -0.12$), Pt / Co ($\theta_\mathrm{SH} = 0.07$) and $\beta$-W / CoFeB ($\theta_\mathrm{SH} = -0.4$).\cite{Liu2012a, Liu2012b, Avci2012, Pai2012, Hao2015a}

In recent works the role of the interfaces of the ferromagnetic film was studied.\cite{Pai2014, Zhang2015, Nguyen2015, Li2017} The interfaces are decisive for the perpendicular magnetic anisotropy,\cite{Pai2014} interlayers between the heavy metal and the ferromagnet can significantly alter the spin transmission and thereby lead to a change of the observed spin Hall angle,\cite{Zhang2015, Nguyen2015} and ferromagnetic layers decorated with ultrathin C layers were shown to have significantly modified perpendicular magnetic anisotropy (PMA) and effective spin Hall angle.

In the present work we study the influence of submonolayer noble metal films inserted between a Ta film and an ultrathin CoFeB layer with PMA. We demonstrate that the PMA and the coercive fields depend sensitively on the choice of interlayer material and that the switching current density is greatly reduced with respect to the reference system without an interlayer. Most remarkably, we find that despite the spin Hall angle of our Ta film is comparatively small, small switching current densities are observed, which we ascribe to the very low PMA or the small coercive field of the ferromagnetic layer with a noble metal interlayer.

\begin{figure}[b]
\includegraphics[width=8.6cm]{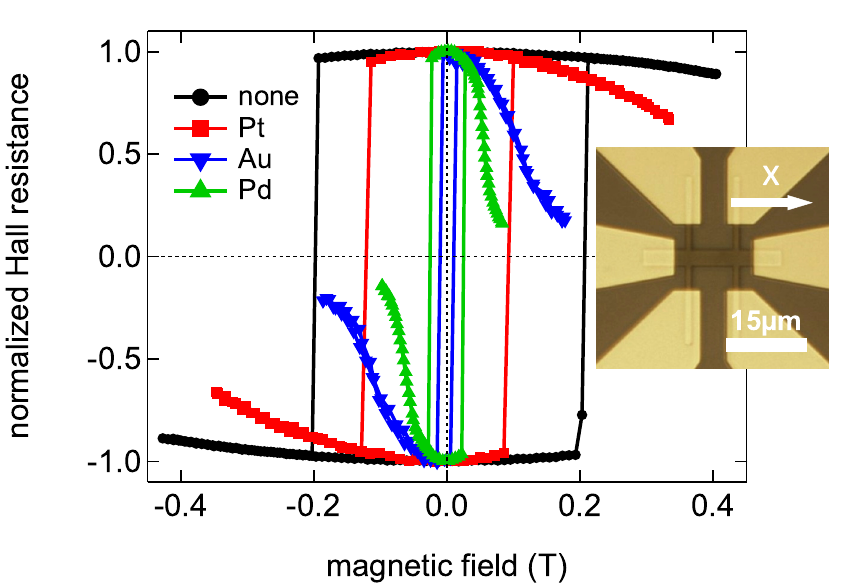}
\caption{\label{fig:AHE_inplane} Normalized Hall resistance as a function of the in-plane magnetic field $B_x$ for samples without (none) and with 2\,\AA{} noble metal interlayers. The inset displays an optical micrograph of a typical Hall bar structure used in this study.}
\end{figure}

Thin films of the type Si substrate / SiO$_x$ 50\,nm / Ta 6\,nm / NM $0.2\,$nm / CoFeB 0.7\,nm / MgO 1.7\,nm / Ta 1.5\,nm (noble metal NM = Pt, Pd, Au) were prepared by dc magnetron sputtering in a 4-inch sputtering system with a base pressure of $p_0 = 2 \times 10^{-7}$\,mbar. The Argon working pressure was set to $1.3 \times 10^{-3}$\,mbar for the metallic layers and $4 \times 10^{-2}$\,mbar for the MgO layer. The films were annealed in vacuum at 275$^\circ$C for 60\,min (160$^\circ$C without an interlayer) and patterned by electron beam lithography and ion beam milling into Hall bar structures with dimensions of $3 \times 15$\,$\mu$m$^2$ (see inset in Figure \ref{fig:AHE_inplane}). The average resistivity of the film stacks was $210\,\mu\Omega\mathrm{cm}$ and the resistance of the current line of the Hall bar structures was reproducibly $(1580 \pm 30)\,\Omega$, with little variation upon interlayer insertion.

\begin{figure}[b]
\includegraphics[width=8.6cm]{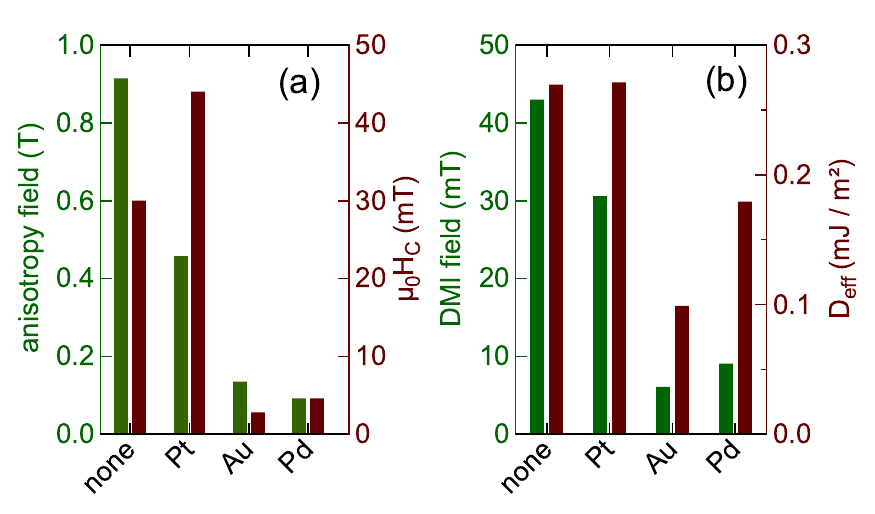}
\caption{\label{fig:extracted} a): Extracted anisotropy fields from fits of equation \ref{eq:anisotropy} to data in Figure \ref{fig:AHE_inplane} and coercive fields of the same Hall crosses on the same samples. b): DMI fields obtained from fits of equation \ref{eq:jc_vs_Bx} to data in Figure \ref{fig:jc_vs_Bx} and effective DMI constants as discussed in the main text.}
\end{figure}

We begin our data analysis by evaluating the anomalous Hall effect (AHE) as a function of the external magnetic field to determine the coercive fields and the (effective) anisotropy fields. In a first step, the coercive fields were determined along the easy axis. To determine the anisotropy fields, in-plane magnetic field loops were recorded with the sample slightly canted against the magnetic field to avoid multidomain configurations. In both cases, the AHE was probed with a current small enough to keep the influence of the spin-orbit torques negligible ($< 5 \times 10^9$\,A/m$^2$). The in-plane magnetic field loops are shown in Figure \ref{fig:AHE_inplane}. A strong dependence on the material of the ultrathin interlayer is clearly observed: all interlayers lead to a significant reduction of the anisotropy field, which is most pronounced for Au and Pd. To quantify the anisotropy fields we fit the standard expression
\begin{equation}\label{eq:anisotropy}
\frac{R_{xy}(B_x)}{R_{xy}(0)} = \cos\left( \sin^{-1}\frac{B_x}{B_\mathrm{an}} \right)
\end{equation}
to the normalized AHE loops. The resulting values of $B_\mathrm{an}$ are summarized together with the corresponding coercive fields in Figure \ref{fig:extracted}\,(a). Notably, the Pt interlayer leads to an increase of the coercive field, whereas the Au and Pd interlayers lead to a strongly reduced coercive field of the CoFeB layer. The anisotropy field is reduced by an order of magnitude with the Au and Pd interlayers. 

\begin{figure}[t]
\includegraphics[width=8.6cm]{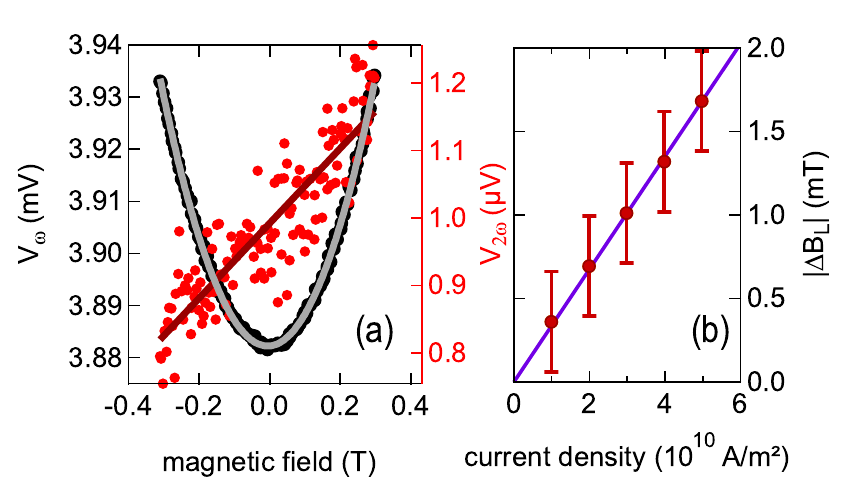}
\caption{\label{fig:harmonic_Hall} a) Harmonic Hall voltages obtained from a magnetic field-sweep in the sample plane at a current density of $3.8 \times 10^{10}\,\mathrm{A/m^2}$. A sample without an interlayer was measured here. Full lines correspond to parabolic and linear fits, respectively. b): Longitudinal effective field as a function of current density as obtained from equation \ref{eq:harmonic_Hall}.}
\end{figure}

The antidamping-like, longitudinal effective fields were measured using the harmonic Hall voltage method with a lock-in amplifier at a frequency of 1031\,Hz. In the regime of $B_x \ll B_\mathrm{an}$ one can evaluate the longitudinal effective field as\cite{Hayashi2014}
\begin{equation}\label{eq:harmonic_Hall}
\Delta B_\mathrm{L} = -2 \left( \frac{\partial V_{2\omega}}{\partial B_x} \bigg/ \frac{\partial^2 V_{\omega}}{\partial B_x^2} \right),
\end{equation}
with the first and second harmonic anomalous Hall voltages $V_\omega$ and $V_\mathrm{2\omega}$. Specifically, we simultaneously measure the in-phase component of $V_\omega$ and the out-of-phase component of $V_{2\omega}$.\cite{Hayashi2014} Here, we neglect contributions from the planar Hall effect and field-like, transverse spin-orbit torques. In Figure \ref{fig:harmonic_Hall}\,(a) we exemplarily show a measurement for the reference sample with no interlayer. The derivatives are obtained by fitting the measured data with a parabola and a line, respectively, so that the effective field can be written as a simple ratio of the fit parameters. In Figure \ref{fig:harmonic_Hall}\,(b) we demonstrate that the observed longitudinal torque is directly proportional to the driving current density. We find a spin-orbit torque efficiency of $\chi_\mathrm{L} = \Delta B_\mathrm{L} / j = 3.36\,\mathrm{mT}/(10^{11}\,\mathrm{A/m^2})$. It is assumed that the current flow is homogeneous within all metallic layers of the stacks and that the Ta cap layer is completely oxidized. The effective spin Hall angle can be estimated by $\theta_\mathrm{SH}^\mathrm{eff} = \frac{2e}{\hbar} \chi_\mathrm{L} M_\mathrm{s} t_\mathrm{F}$ with the saturation magnetization $M_\mathrm{s}$ and the thickness $t_\mathrm{F}$ of the ferromagnetic layer. The CoFeB used in our stacks has a saturation magnetization of typically $M_\mathrm{s} \approx (1000 \pm 100)\,\mathrm{kA/m}$, so we obtain $\theta_\mathrm{SH}^\mathrm{eff} \approx -0.07 \pm 0.01$, which is somewhat smaller than usually reported values for Ta.\cite{Hao2015b} With the interlayers, we consistently find a smaller spin Hall angle of $\theta_\mathrm{SH}^\mathrm{eff} \approx -0.05$ in all cases. This can be explained by a reduced interface transparency for the spin current,\cite{Zhang2015} or by the positive intrinsic spin Hall angle of the three interlayer materials and is therefore consistent with expectation.

\begin{figure}[b]
\includegraphics[width=8.6cm]{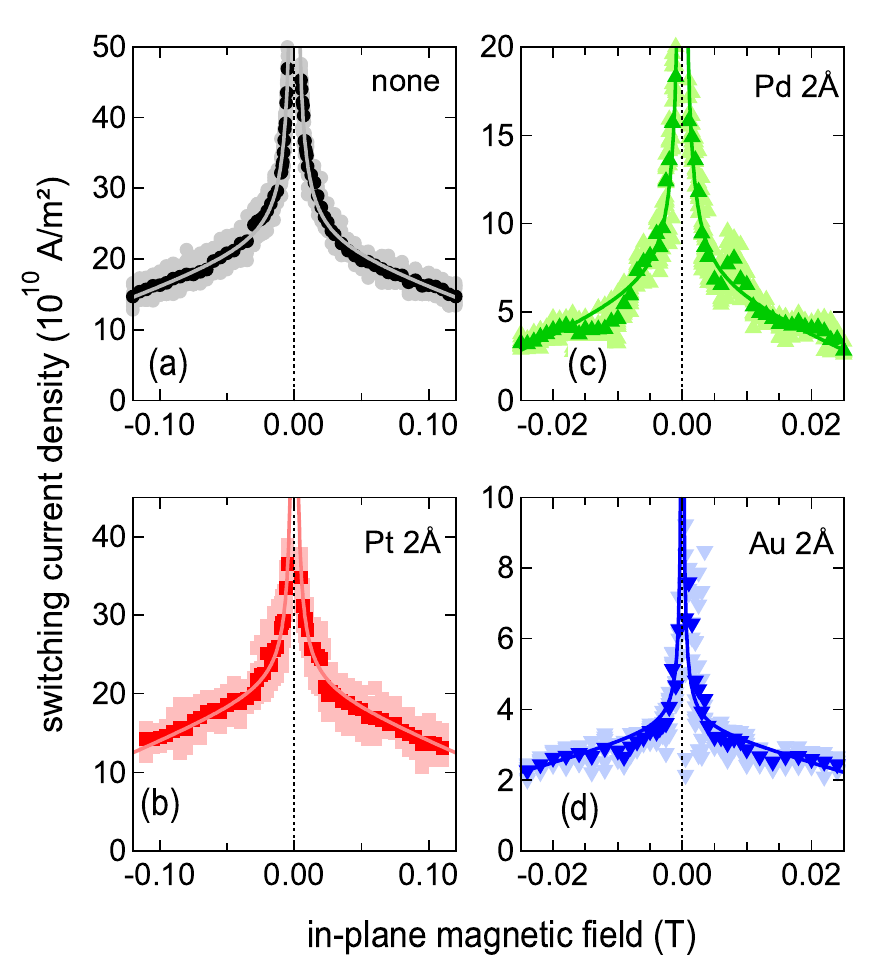}
\caption{\label{fig:jc_vs_Bx} Switching current density $j_c$ as a function of in-plane magnetic field for the sample without interlayer (a) and Pt (b), Pd (c), Au (d) interlayers. Light colored markers are from individial pulsed current loops,  dark colored markers are averages over six repeats. Full lines are fits of equation \ref{eq:jc_vs_Bx} to the averaged data. Note the different scales in the figures.}
\end{figure}

Current-induced magnetization switching (CIMS) experiments were performed with pulsed currents driven by a two-channel arbitrary waveform generator (Agilent 33522A) in an internally synchronized, differential mode. The circuit impedance allowed to obtain well-defined voltage pulses down to a pulse-width of $\tau = 100\,\mathrm{ns}$ and $U = 20\,\mathrm{V}$. To obtain the dependence of the switching current density $j_\mathrm{c}$ as a function of the in-plane magnetic field, single pulses with $\tau = 500\,\mathrm{ns}$ were injected into the Hall bars. The amplitude was swept from negative to positive and back to negative to obtain a pulsed current loop. Between the pulses, the AHE was measured with a small probing current to determine the magnetization state. For each value of the external magnetic field, six pulsed current loops were recorded for estimating an average switching current density at the given field value from an automatic fitting procedure. The results from the field-dependent CIMS experiments are given in Figure \ref{fig:jc_vs_Bx}\,(a)-(d). In all cases we observe the typical, symmetric behaviour with a steep increase of the switching current density at small $B_x$.\cite{Hao2015a, Hao2015b} Notably, the four samples fall into two groups: without the interlayer or with the Pt interlayer, large switching current densities of about $1.5\times 10^{11}\,\mathrm{A/m^2}$ at $B_x = 0.1\,\mathrm{T}$ are obtained and rather large magnetic fields are necessary to reach the linearly falling regime. With the Pd and Au interlayers, much smaller switching current densities as low as $2.5 \times 10^{10}\,\mathrm{A/m^2}$ at $B_x = 0.02\,\mathrm{T}$ are obtained, similar to a previous report of native Ta / CoFeB / MgO systems with thicker CoFeB and thereby lower anisotropy as compared to our case.\cite{Hao2015b} Here, the low switching currents are obviously related to the small anisotropy and coercive fields in these samples, although we do not find a strict one-to-one correspondence between any two of these quantities.

The strongly enhanced switching current density at low $B_x$ is interpreted as an effect of the Dzyaloshinskii-Moriya interaction, which gives rise to N\'eel-type domain walls of a single chirality and needs to be broken by an external magnetic field to allow the domains to expand and thereby switch the magnetization.\cite{Perez2014, Yu2014, Pai2016} The field at which the switching current becomes constant is then identified as an effective DMI field $B_\mathrm{DMI}$. In our measurements we do, however, not find the idealized behaviour of two intersecting line segments,\cite{Hao2015a} so we need to find a procedure to unambigously determine the DMI field. We propose to fit the data with an empirical expression of the type
\begin{equation}\label{eq:jc_vs_Bx}
j_\mathrm{c}(B_x) =  a |B_x| + b |B_x|^n + C,
\end{equation}
where $a < 0$, $b,C > 0$, and $n < 0$. We find very good agreement with the data for $n \approx -1$ as is shown in Figure \ref{fig:jc_vs_Bx}, so to reduce the number of fit parameters we keep $n=-1$ fixed. To identify the DMI field, we find the maximum of the curvature $\kappa = j_\mathrm{c}''/(\sqrt{1 + (j_\mathrm{c}')^2)}^{3}$. The resulting DMI fields are summarized in Figure \ref{fig:extracted}\,(b). One finds, that the DMI fields follow the same trend as the anisotropy fields for the various interlayers. The DMI constant can now be obtained as $|D_\mathrm{eff}| = M_\mathrm{s} B_\mathrm{DMI} \sqrt{A/K_\mathrm{eff}}$ with the anisotropy constant $K_\mathrm{eff} = M_\mathrm{s} B_\mathrm{an} / 2$ and the spin-stiffness constant $A = 18 \times 10^{-12}\,\mathrm{J/m}$ is assumed. The resulting values of $D_\mathrm{eff}$ are shown in Figure \ref{fig:extracted}\,(b). Without an interlayer, we obtain $D_\mathrm{eff} \approx 0.27\,\mathrm{mJ/m^2}$, which is close to previously reported values on Ta / CoFeB / MgO systems.\cite{Torrejon2014, Pai2016} This supports both the adequacy of our procedure to extract the DMI field as well as the correctness of the interpretation of the enhanced switching current density at small magnetic field as an effect of the DMI. The Pt interlayer does not modify the DMI significantly, whereas both the Pd and Au interlayers give rise to a strong reduction of the DMI. For Pt, one would naively expect a strong enhancement of the DMI, however it was shown that the DMI depends on the thickness of the heavy metal layer and, specifcally for the case of Pt, our value of $D$ is in agreement with a recent study on the thickness dependence on Pt / CoFeB films, where saturation at 2\,nm Pt thickness with $D \approx 0.45\,\mathrm{mJ/m^2}$ is found.\cite{Tacchi2017} In all cases, the ratio of the DMI field and the anisotropy field is way below the limit for Skyrmion crystals or Dzyaloshinskii spiral states ($B_\mathrm{DMI} / B_\mathrm{an} = D_\mathrm{eff} / \sqrt{4K_\mathrm{eff}A} \geq 2 / \pi$).\cite{Heide2008} The very small DMI field observed for the Au and Pd interlayers is however beneficial for field-free CIMS, where an effective in-plane field is obtained from exchange bias or interlayer exchange coupling.\cite{Fukami2016, Lau2016}

\begin{figure}[t]
\includegraphics[width=8.6cm]{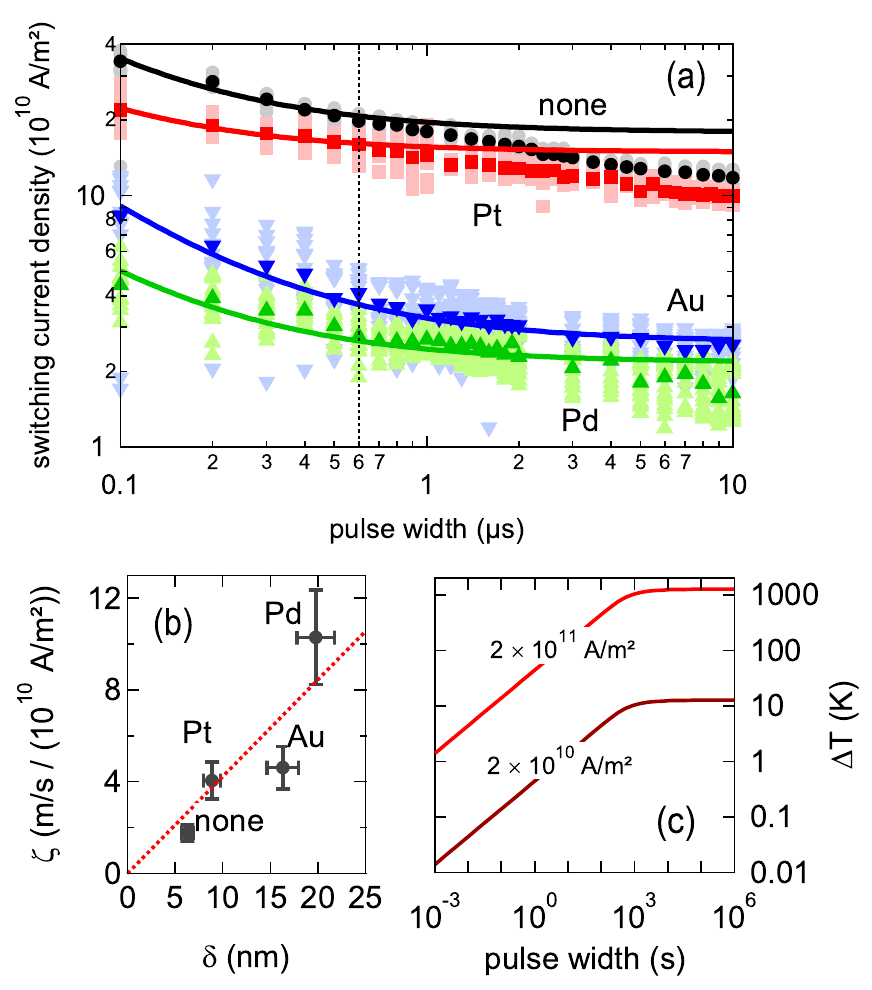}
\caption{\label{fig:jc_vs_tau} (a): Switching current density $j_\mathrm{c}$ as a function of the pulse width for the different samples. Light colored markers are from individial pulsed current loops, dark colored markers are averages over ten repeats. Full lines are fits to the averaged data as discussed in the main text. The dashed line represents the upper limit of the fitting range.(b): Domain wall mobility $\zeta$ as defined in the main text as a function of the calculated domain wall width $\delta = \sqrt{A/K_\mathrm{eff}}$. (c): Calculated temperature of the current line as a function of the pulse width according to the model of Ref. \onlinecite{You2006}, assuming an SiO$_2$ substrate as a worst-case estimate. }
\end{figure}

Finally, we investigate the pulse width dependence of the switching current density for the four interlayers. For each pulse width, ten pulsed current loops were recorded to estimate the average switching current density at the given pulse width. An in-plane field slightly larger than the corresponding DMI field was applied parallel to the current channel to align the domain walls (DW). In Figure \ref{fig:jc_vs_tau}\,(a) we show the experimental results together with fits to the expression $j_\mathrm{c}(\tau) = j_{c,0} + a /\tau$. Here, $j_{c,0}$ is the threshold current for DW nucleation or depinning (whichever is larger) The DW propagates with a velocity $v_\mathrm{DW}$ proportional to the current density; for a shorter pulse, a higher velocity and therefore a higher current density are required to switch the magnetization within the Hall bar. According to the DW propagation picture, $a$ can be written as $a = w/\zeta$, where $v_\mathrm{DW} = \zeta j$ defines the mobility with respect to the current density $\zeta$ and $w$ is the length scale of the Hall bar device, here taken as its width. We obtain $a \approx 2900 \dots 17000\,\mathrm{As/m^2}$ for the different interlayers. The meaning of the parameter is clear: it corresponds to the amount of charge  that needs to be passed beneath a unit area of the ferromagnetic layer (and thereby angular momentum via the spin Hall effect injected into the ferromagnetic layer) to switch the magnetization. In the simplest theoretical model, DW velocity is expressed as $v_\mathrm{DW} = \frac{\gamma}{\alpha} B \sqrt{A/K_\mathrm{eff}}$, neglecting DW pinning.\cite{Beach2005} Here, $\gamma$ is the gyromagnetic ratio, $\alpha$ is the Gilbert damping and the magnetic field $B$ has to be substituted by the longitudinal effective field from the spin Hall effect $\Delta B_\mathrm{L}$. Therefore, one may expect a linear scaling of the domain wall mobility $\zeta$ with the domain wall width $\delta = \sqrt{A/K_\mathrm{eff}}$. As shown in Figure \ref{fig:jc_vs_tau}\,(b), we find that this scaling is fulfilled within the uncertainties due to the fitting procedure and the saturation magnetization. Substituting the effective field $\Delta B_\mathrm{L} = \chi j$ into $v_\mathrm{DW}$ one can express the domain wall mobility as $\zeta = \gamma \chi \delta / \alpha$. By fitting a line in Figure \ref{fig:jc_vs_tau}\,(b) one obtains $\zeta / \delta = 0.042\,\mathrm{m^2/As}$ and assuming the gyromagnetic ratio to be that of the free electron, we calculate $\alpha = 0.14$, a realistic value for ultrathin CoFeB.\cite{Malinowski2009} Thus, the pulse width dependence of the switching current density is compatible with the DW nucleation and propagation picture of the spin Hall driven CIMS.\cite{Yu2014} For the samples without interlayer or with a Pt interlayer, the negative slope at long pulses is easily explained by thermally activated nucleation/depinning due to the Joule heating. Since $\Delta T \propto \rho j^2 \tau$ for sufficiently short pulses ($\ll 1\,\mathrm{ms}$, cf. Figure \ref{fig:jc_vs_tau}\,(c)), it is obvious that the temperature increase is about two orders of magnitude smaller in the sample with the Au or Pd interlayers as compared to the reference sample, for which thermal activation considerably reduces the switching current. By making use of the 2D model of the current line temperature derived by You et al. we estimate the temperature increase to be about 200\,K at $2 \times 10^{11}\,\mathrm{A/m^2}$ and only 2\,K at $2 \times 10^{10}\,\mathrm{A/m^2}$ after a 10\,$\mu$s pulse, cf. Figure \ref{fig:jc_vs_tau}\,(c).\cite{You2006} With pulse widths of a few hundred nanoseconds, the temperature increase will be below 50\,K at $j = 2 \times 10^{11}\,\mathrm{A/m^2}$, so that thermal activation by Joule heating will play only a minor role for short pulses.\cite{Neumann2016}

To summarize, we demonstrate that the magnetic properties of a Ta / CoFeB / MgO heterostructure can be tuned by noble metal interlayer insertion without loss of the longitudinal effective field to the spin Hall effect. As a consequence of reduced anisotropy, the Dzyaloshinskii-Moriya effective field is greatly reduced, making field-free switching easily achievable. Low switching current densities of the order of $2 \times 10^{10}\mathrm{A/m^2}$ are obtained with 2\,\AA{} Au or Pd interlayers. An analysis of the pulse width dependence of the switching current density allows for an estimate of the domain wall velocities and demonstrates that the magnetization is switched by expanding domains.

We thank Daniel Meier for help with the sample preparation and G\"unter Reiss for making available the laboratory equipment. M.M. further thanks Can Onur Avci and Manuel Baumgartner for fruitful discussions.

\end{document}